\documentstyle[prd,aps,preprint]{revtex}
        		
\newcommand\ee{\end{equation}}
\newcommand\be{\begin{equation}}
\newcommand\eea{\end{eqnarray}}
\newcommand\bea{\begin{eqnarray}}

\newcommand\Mpl{M_{\rm Pl}}

\newcommand\lsim{\mathrel{\rlap{\lower4pt\hbox{\hskip1pt$\sim$}}
    \raise1pt\hbox{$<$}}}
\newcommand\gsim{\mathrel{\rlap{\lower4pt\hbox{\hskip1pt$\sim$}}
    \raise1pt\hbox{$>$}}}

\def\dslash{\not{\hbox{\kern-2pt $\partial$}}}
\def\Dslash{\not{\hbox{\kern-4pt $D$}}}
\def\Oslash{\not{\hbox{\kern-4pt $O$}}}
\def\Qslash{\not{\hbox{\kern-4pt $Q$}}}
\def\pslash{\not{\hbox{\kern-2.3pt $p$}}}
\def\kslash{\not{\hbox{\kern-2.3pt $k$}}}
\def\qslash{\not{\hbox{\kern-2.3pt $q$}}}
 \newtoks\slashfraction
 \slashfraction={.13}
 \def\slash#1{\setbox0\hbox{$ #1 $}
 \setbox0\hbox to \the\slashfraction\wd0{\hss \box0}/\box0 }
 
\def\eeq{\end{equation}}
\def\beq{\begin{equation}}

\begin{document}

\preprint{FERMILAB-PUB-97/276-A}
\draft
\tighten

\title{Cosmological Implications of Low Energy \\
Supersymmetry Breaking Models}
\author{Antonio Riotto\footnote{Email: riotto@fnas01.fnal.gov.}}
\address{NASA/Fermilab Astrophysics Center,\\ Fermilab
National Accelerator Laboratory, 
Batavia, Illinois~~60510.
}
\date{August 1997}
\maketitle

\begin{abstract}
We show that stable  local cosmic strings are a generic prediction of 
supersymmetric models where supersymmetry is broken at scales  within a few orders of magnitude of the weak scale and is fed down to the observable sector by gauge interactions. The typical energy scale of these strings is about $10^2$ TeV. 
Since in the  most general  and natural  scenario   there are two different sources of contributions to ordinary superpartner masses,  gauge mediation from the messanger sector and  anomalous $U(1)$ $D$-terms,    novel features    are that  the cosmic strings may be superconducting with the role of bosonic charge carriers played by  squarks and sleptons and that   some baryonic or leptonic charge may be stored in the core of the string. Possible cosmological and astrophysical   implications of these superconducting strings  are discussed.

\end{abstract}

\newpage
\setcounter{page}{1}
\baselineskip=20pt

{\bf 1.}~~Supersymmetry (SUSY) provides solutions  to solve many of the puzzles of the
standard model such as the stability of the weak scale under radiative
corrections as well as the origin of the weak scale itself. Local
supersymmetry provides a promising way to include gravity within the
framework of unified theories of particle physics,  eventually leading the
way  to a theory of everything in string models. For these very good
reasons, supersymmetric extensions of the standard model (MSSM) 
have been the focus of intense theoretical activity  in recent
years \cite{susy}. Since experimental observations require supersymmetry to
be broken, it is essential to have a knowledge of
the nature and the scale of supersymmetry breaking in order to have a
complete understanding of the physical implications of these theories.
At the moment, we lack such an understanding and therefore it is important 
to explore the various ways in which supersymmetry breaking can arise and
study their consequences. 

There are several hints from the study of general
class of MSSM which could perhaps be useful in trying to explore
the nature of supersymmetry breaking. 
Two particular ones that rely on the supersymmetric sector of model are {\it (i)} 
the natural suppression of flavor changing neutral
currents (FCNC),   which require either a high degree of degeneracy among squarks
of different flavor or 
the first two generations of squarks to be  much heavier than the weak scale,  and {\it (ii)} stringent upper limits on the electric
dipole moment of the neutron (EDMN),  which imply severe constraints on the
gaugino masses as well as on the $A$ and $B$ terms of MSSM \cite{dine}.
One could take the point of view that these observational facts may be telling
us something about the nature of supersymmetry breaking. If this is true,
then it is important to isolate those SUSY breaking scenarios which 
realize the above properties in a simple manner and study their implications.

 The most  common and popular  approach is
to implement supersymmetry breaking in some  hidden
sector where some $F$-term gets a vacuum expectation value (VEV)  and then transmit it to the standard model sector  by
gravitational interactions. This is the so-called hidden $N=1$ supergravity scenario \cite{susy}. If one arranges the parameters in the hidden sector in such a way that the typical $\langle F\rangle $-term is of the order of $\langle F\rangle^{1/2} \sim\sqrt{(1\:{\rm TeV}) \Mpl}\sim 10^{11}$ GeV, where $\Mpl=2.4\times 10^{18}$ GeV,   the gravitino mass $m_{3/2}$ turns out to be of the order of  the TeV scale. In the $N=1$ supergravity scenario, however, the flavor-blindness is likely to be spoiled in the K\"{a}hler potential. One possibility is that the underlying theory possesses some flavor symmetry, but these symmetries must be broken and it is hard to reconcile the strong violation of flavor in the fermion sector with the degeneracy in the sfermion sector. The suppression of the EDMN is also hard to explain in supergravity since  one expects the bilinear and trilinear terms to be of the order of $m_{3/2}$ and explicit CP-violating phases to be ${\cal O}(1)$.    

In this paper we will concern ourselves with  alternative   scenarios where  supersymmetry is broken  dynamically or via the O'Raifeartaigh mechanism at low energies, within a few orders of magnitude of the weak scale.   
 
Gauge interactions can serve as the messangers of supersymmetry breaking 
and the SUSY breaking mediators are fields that transform nontrivially under
the  standard model gauge group \cite{gmsb}. These are the so-called gauge mediated supersymmetry breaking (GMSB) models. The soft SUSY breaking mass terms of sfermions get  contributions at two-loop,   $\widetilde{m}^2\sim  
\left({{\alpha}\over{4\pi}}\right)^2\Lambda^2$, where $\Lambda\sim 10$ TeV  is the typical  scale of the messanger sector. The GMSB models have the extra advantage that
the FCNC effects are naturally suppressed. This is due to the fact that   
 at the scale $\Lambda$ the squark masses are
all degenerate because of  the flavor blindness of the standard model gauge group. Only a slight asymmetry is introduced by renormalization group extrapolation from the scale $\Lambda$ to low energies. Moreover, the trilinear soft breaking terms $A$ vanish at $\Lambda$ rendering the CP-violation problem
milder. 

Recently,  an alternative approach have been proposed in which gauginos, higgsinos and the third generation of squarks are sufficiently light to stabilize the electroweak scale, but the two first generations of squarks and sleptons  are sufficiently heavy to suppress the FCNC and the EDMN below the experimental bound \cite{pomarol,dudas,more}. This class of models is dubbed "more" minimally supersymmetric than the MSSM since they do not require some ad hoc supposition of degeneracy or alignment \cite{more}. In most of the attempts made so far along this line \cite{pomarol,dudas,riotto,nelson} 
the  crucial feature  is the
existence of an anomaly-free $U(1)$ gauge group, which appears anomalous below  some scale $M$. The effective theory includes a  Fayet-Iliopoulos (FI) $D$-term proportional to ${\rm Tr}\:Q M^2$ \cite{w}, where the  $Q$'s represent the $U(1)$ charges of the heavy fields wich have been integrated out\footnote{
In this paper we will not be concerned with the case in which the $U(1)$ group emerges from string theories and  
the  anomaly is cancelled by the Green-Schwarz
mechanism. The FI $D$-term is expected to be of the order of the 
stringy scale  $(10^{17}-10^{18})$ GeV.  Cosmic strings in the framework of anomalous $U(1)$ superstring theories have been discussed in \cite{casas} and recently in \cite{lr}. The mass of the  $U(1)$ gauge field  receives contribution from the dilaton VEV $\langle S+\overline{S}\rangle$ and from any other field $\Phi$
which is charged under $U(1)$ and has  nonvanishing VEV.  Since the dilaton field, contrary to the $\Phi$ field,  is not winding at infinity, the gauge flux will be non-integer. This means that   strings arising from  anomalous $U(1)$ superstring theories are not local, but semiglobal \cite{gd}. 
}. If the first two generations of squarks and sleptons, contrary to the third generation,  carry  $U(1)$ 
charges, the required mass hierarchy is obtained. 

The suppression of  the FCNC and the EDMN are not the only good features
a supersymmetric model should possess though. 
SUSY models can be regarded as  realistic  only when they are   able to reproduce the quantitative features of the fermion spectrum and the CKM matrix. In a unified picture it is  desirable that the solution to all these puzzles  reside in the same sector of the theory. Recent investigations   have focused on the possibility that  a Frogatt-Nielsen flavor mechanism \cite{fn} is implemented
in the messanger sector \cite{riotto,nelson}. This means that the  messenger sector is also to be a Frogatt-Nielsen sector and that the 
messenger $U(1)$ symmetry by which the two first generation fermions get large masses is the same  Frogatt-Nielsen $U(1)$ for quarks and leptons.
Therefore, the most general (and maybe natural) scenario is an hybrid one  where  there are two different sources of contributions to ordinary superpartner masses,  gauge mediation from the messanger sector and  anomalous $U(1)$ $D$-terms \cite{riotto,nelson,inprep}.

The cosmological aspects of the   class of supersymmetric models 
in which SUSY breaking takes place at low energies, of the order of $10^3$ TeV or so,   has only been partially explored. It has been suggested  that viable cold dark matter candidates may be  present in the messanger sector \cite{gian}.  Inflation might    occur  at late stages of the evolution of the Universe when its  energy density  is dominated by a  vacuum energy density    related to
 the same $F$-term responsible for the spectrum of supersymmetric particles in
GMSB models \cite{dr,dr2}. 

In this paper we consider the question of whether one might expect the presence of stable cosmic strings in low energy SUSY breaking models. Over the last two decades, cosmic strings have evoked a great deal of interest  as possible remnants of a Grand Unified era in the early Universe as well as possible mechanism for structure formation \cite{vil}. 
However, no compelling particle physics model exist that give rise to such defects. 

We would like to  show that low energy SUSY breaking models are good candidates. This does not come as a surprise because of the rich gauge group structure. The novel feature    is that these cosmic strings may be superconducting with the role of bosonic charge carriers played by  squarks and sleptons. This means that in the core of the strings not only the electromagnetic gauge group is spontaneosly broken, but also  the global  baryon ($B$) and the lepton ($L$) symmetries are broken.  

This kind of  superconducting cosmic strings may have   important  cosmological and astrophysical 
effects. Since sfermions may carry a baryonic $U(1)$ global charge, a certain amount of baryonic charge may be present inside the strings. As a consequence, 
the  baryon asymmetry might be generated in some  unconventional way.  
Primordial
 magnetic fields might be formed by a network of charged-current carrying cosmic strings. Such fields could
 account for the observed galactic and intergalactic magnetic fields since they might be suffice to seed magnetic dynamos on galactic scales. We will comment about these and other cosmological aspects in the following.  
Let us first show that cosmic strings are a generic prediction of low energy SUSY breaking models. 

{\bf 2.}~~As  is known, stable cosmic strings arise when the manifold ${\cal M}$ of degenerate vacua has a non-trivial first homotopy group, $\Pi_1({\cal M})\neq {\bf 1}$. A generic feature of low energy SUSY breaking models is that the gauge group of the secluded sector,  where supersymmetry  is broken, may be  of  the form ${\cal G}\otimes U(1)_m$. The group $U(1)_m$ is usually some global $U(1)$  which is gauged and made anomaly-free.   
The fields  in the messanger sector  are charged under $U(1)_m$ and (some of them) under the standard model gauge group.   When supersymmetry is broken in the secluded sector,  some scalar fields in this sector  may acquire a vacuum expectation value (VEV), but leave  the $U(1)_m$ gauge symmetry unbroken.  
As a toy model, 
let us consider the following superpotential in the secluded sector where a gauge $U(1)_m$ symmetry has been imposed and $X$ is a gauge singlet
\begin{equation}
W=\lambda X \left({\bar \Phi}_1 \Phi_1 - m^2 \right)
+M_1 {\bar \Phi}_1 \Phi_2 + M_2 {\bar \Phi}_2 \Phi_1 ~.
\label{orafeq}
\end{equation}
 For $\lambda^2 m^2 \ll M_1^2,M_2^2$, the vacuum of this model is such
that $\langle \phi_i \rangle =\langle {\bar \phi}_i \rangle =0$
($i=1,2$), where  $\bar{\phi}_i$ and $\phi_i$ are the scalar components of the superfields $\bar{\Phi}_i$ and $\Phi_i$,  respectively. Supersymmetry is broken by the O'Raifeartaigh mechanism
and $\langle F_X \rangle =-\lambda^2 m^2$. The $U(1)$ gauge group remains unbroken at this level. Because of supersymmetry breaking,  a term like $V=(F_X{\bar \phi}_1 \phi_1+{\rm h.c.})$ will appear
 in the potential. It is easy to show that, integrating out the $\phi_{i}$ and $\bar{\phi}_i$ scalar fields, induces a  nonvanishing FI  $D$-term 
\begin{equation}
\xi\sim \frac{F_X^2}{16\pi^2(M_1^2-M_2^2)}\ln\left(\frac{M_2^2}{M_1^2}\right). 
\end{equation}
The presence of the anomalous FI $D$-term may induce the spontaneous breakdown of the residual   $U(1)_m$ gauge symmetry  along some field direction in the messanger sector. In this case 
 local cosmic strings are formed. Indeed, let us make the very general assumption that integrating out the heavy fields belonging to the secluded sector amounts to generate a positive  one-loop  FI $D$-term $\xi$  and 
 negative two-loop soft SUSY breaking squared masses  for (some of)  the scalar fields $\phi_i$ of the messanger sector. These squared masses are  generally proportional to the square of the corresponding  $U(1)_m$ charges $q_i$.  The
potential will read
\begin{equation}
\label{pot}
V=-\sum_iq_i^2\widetilde{m}^2\left|\phi_i\right|^2+\sum_i\left|\frac{\partial W}{\partial\phi_i}\right|^2+\frac{g^2}{2}\left(\sum_i q_i\left|\phi_i\right|^2+\xi\right)^2,
\end{equation}
where $g$ is the $U(1)_m$ gauge coupling constant.  From the minimization of (\ref{pot}), one can see that among the fields with $F$-flat directions, the one with the smallest negative charge (call it $\varphi$) will get a VEV\footnote{This particular field cannot be charged under the standard model gauge group, but may  play the role of the  Frogatt-Nielsen field to generate appropriate fermion mass matrices \cite{nelson,inprep}.}
\begin{equation}
\langle|\varphi|^2\rangle=\frac{1}{q_\varphi}\left(q_\varphi\frac{\widetilde{m}^2}{g^2}-\xi\right).
\end{equation}
When this happens, the  residual global  $U(1)_m$ symmetry gets broken leading to the formation of   local strings. 

The string mass per-unit-length is given by $\mu\sim\xi$. Since $\sqrt{\xi}$ is a few orders of magnitude larger than the weak scale, 
cosmic strings are not very heavy (at least in comparison with the most commonly discussed Grand Unified strings) and they are expected to play no role in the generation of the temperature anisotropies in the cosmic microwave background. Yet, they may have other cosmological consequences.  

If there are extra local or global $U(1)$ symmetries in the messanger sector, they   can be also spontaneously broken by the FI anomalous $D$-term. This happens necessarily if there are no singlet fields charged under the anomalous gauge  $U(1)_m$ only. In such a case, besides the previously mentioned  cosmic strings, there may arise cosmic strings associated to the breaking of extra $U(1)$ factors. However,  the condition to produce cosmic strings is $\Pi_1({\cal M})\neq {\bf 1}$ and one must consider the structure of the {\it whole} potential, {\it i.e.} all the $F$-terms and all the $D$-terms. When this is done, it turns out that, depending on the specific models, some (or all) of the global and local cosmic strings may disappear. In general,  there can be models that have just global strings, just local strings, or both global and local strings. Thus, the analysis must be done case by case, but 
it is important to remember  that the class of models predicting   global strings are astrophysically ruled out. Indeed, the Goldstone particles associated to the spontaneous breakdown of the global $U(1)$  are expected to have interactions with the ordinary matter suppressed by the same mass scale of the global cosmic string. Being  $\sqrt{\xi}$ only a few orders of magnitude
larger than the weak scale, these Goldstone particles would be copiously emitted from supernovae and stars at an   unacceptable level \cite{raffelt}. 
This, in particular, may exclude those models in which the secluded and the messanger sectors communicate only through the  gauged $U(1)_m$ interaction and  no renormalizable couplings  among the superfields of the two sectors  appear in the superpotential 
because of the charge assignment.  When some scalar fields in the secluded sector  acquire a vacuum expectation value  the $U(1)_m$ gauge symmetry may be  spontaneously broken leaving a residual  global $U(1)$ in the messager sector. This residual  global $U(1)$ may be   spontaneously broken by the FI  $D$-term giving rise to 
global cosmic strings and astrophysically harmful  massless Goldstone modes.

{\bf 3.}~~~Let us now show that the cosmic strings produced in the low energy SUSY breaking models are used to be superconducting. Coming back to the potential (\ref{pot}), we suppose that the quark and/or the lepton superfields are charged under the  $U(1)_m$ group.  As we argued, this is in the spirit of the "more" MSSM \cite{more} and somehow welcome when trying to quantitatively predict the fermion mass spectrum and the CKM matrix.  

  One could consider  a flavor-dependent charge assignment; what it is crucial is to equally charge the first and the second family of down-quarks to be consistent with flavor-violating constraints. The charge assignment for the Higgs superfields $H_U$ and $H_D$ is also model dependent; in order to have the top Yukawa coupling unsuppressed, one has usually to require that the up-type Higgs and the left-handed and right-handed top superfields do not carry any $U(1)_m$-charge. If this is the case,  the contribution to the  soft supersymmetry breaking mass terms of the stop sector originate   from gauge interactions with the messanger fields. 
If the down-type Higgs supefield carries a charge, 
$\tan\beta=\langle H_u^0\rangle/ \langle H_d^0\rangle\gg 1$ is a generic prediction \cite{riotto,nelson}. 

Let us focus of one of the sfermion fields\footnote{The role of charge carriers might be played as well by some  scalar fields which are charged under the standard model gauge group and are part of the 
  messanger sector. These fields have zero VEV in the true vacuum and are responsible for gauge mediation.}, $\widetilde{f}$ with $U(1)_m$-charge $q_f$ such that  sign $q_f=$ sign $q_\varphi$. The potential for the fields $\varphi$ and $\widetilde{f}$ is written as
\begin{equation}
V(\widetilde{f},\varphi)=-q_\varphi^2\widetilde{m}^2|\varphi|^2-q_f^2\widetilde{m}^2|\widetilde{f}|^2+
\frac{g^2}{2}\left(q_\varphi |\varphi|^2+q_f|\widetilde{f}|^2+\xi\right)^2+\lambda |\widetilde{f}|^4,
\end{equation}
where we have assumed, for simplicity, that $\widetilde{f}$ is $F$-flat. The parameter $\lambda$  is generated from the standard model gauge group $D$-terms and is vanishing if we take $\widetilde{f}$ to denote a family of fields parametrizing a $D$-flat direction. 

At the global minimum $\langle \widetilde{f}\rangle=0$ and the electric charge, the baryon and/or the lepton numbers are conserved. The soft breaking mass term for the sfermion reads
\begin{equation}
\Delta m_{\widetilde{f}}^2=q_f\left(q_\varphi-q_f\right)\widetilde{m}^2, 
\end{equation} 
which is positive in virtue of the hierarchy $q_\varphi<q_f<0$. Consistency with experimental bound requires $\Delta m^2_{\widetilde{f}}$ to be of the order of $\left(20\:{\rm  TeV}\right)^2$ or so,  which in turn requires $\xi\sim (4\pi/g^2)\widetilde{m}^2\sim (10^2\:{\rm TeV})^2$. Notice that $\Delta m_{\widetilde{f}}^2$ does not depend upon $\xi$. 

Let us analyse what happens in the core of the string. In this region of space, 
the vacuum expectation value of the field vanishes,   $\langle|\varphi|\rangle=0$,  and nonzero values of $\langle |\widetilde{f}|\rangle$ are energetically preferred in the string core 
\begin{equation}
\langle |\widetilde{f}|^2\rangle=\frac{\widetilde{m}^2 q^2_f-g^2\xi q_f}{g^2 q^2_f+2\lambda}.
\end{equation}
Since the vortex is cylindrically symmetric around the $z$-axis, the condensate will be of the form $\widetilde{f}=\widetilde{f}_0(r,\theta)\:{\rm e}^{i\eta_f(z,t)}$ where $r$ and $\theta$ are the polar coordinate in the $(x,y)$-plane.

The above analysis is not sufficient to show the existence of bosonic charge carriers. One must check that the kinetic term for $\widetilde{f}$ also allows a nonzero value of $\widetilde{f}$ in the string. Following  ref. \cite{vacha} we prove that by showing that the equation of motion for $\widetilde{f}$, linearized around $\widetilde{f}=0$, admits growing solutions. This will then show that in the background of the string $\widetilde{f}$ is unstable to the formation of a nonzero condensate on the string. Let us consider the  equation of motion of $\widetilde{f}$
\begin{equation}
\label{sc}
-\partial_\mu\left(\partial^\mu\widetilde{f}-i g q_f R^\mu\:\widetilde{f}\right)=\frac{\partial V(\varphi,\widetilde{f})}{\partial\widetilde{f}^{\dagger}}-g^2 q_f^2 R_\mu R^\mu \widetilde{f},
\end{equation}
where $R_\mu$ is the $U(1)_m$ gauge field which takes the form of the Nielsen-Olesen configuration \cite{no}. We may now linearize Eq. (\ref{sc}) around $\widetilde{f}=0$ and take the following form for the   perturbation 
of $\widetilde{f}$: $ \delta\widetilde{f}={\rm exp}\left(-i\omega t\right)g(r)$. The linearized equation of motion reads
\begin{equation}
-\nabla^2 g(r)+{\cal V}(r) g(r)=\omega^2 g(r),
\end{equation}
where $\nabla^2$ is the two-dimensional Laplacian and ${\cal V}(r)$ represents an "effective" potential. At $r=0$ ${\cal V}= 
-q_f^2\widetilde{m}^2+q_f g^2\xi<0$ and ${\cal V}$ increases monotonically with $r$ until it reaches the asymptotic value $\left[\widetilde{m}^2 q_\varphi\xi-(\widetilde{m}^4 q_\varphi^2/g^2)\right]$. The two-dimensional Schroedinger equation above for $g(r)$ will admit at least one bound state with $\omega^2<0$. Thus $\widetilde{f}$ is unstable to forming a condensate on the string. 

The electric charge, the baryon/and or the lepton numbers are broken inside the string \cite{private}. If there is a condensate $\widetilde{f}$ in the string, then there will massless excitations of the Goldstone boson field $\eta_f$ as well. There are  different  currents on the string world-sheet. One is conserved topologically; 
being a periodic variable, $\eta_f$ may have a net winding number $N$ around a loop of string associated to the corresponding topological current. 
This was shown by Witten 
\cite{superc} to imply the existence of a persistent current. The other  currents are the electromagnetic and the baryon and/or lepton currents, which are  conserved by the equation of motions (at least at the classical level). They may be associated to the relative charges per unit length along the string. In particular, it is intriguing that some baryonic charge may be stored in the core of the string. Indeed, if the sfermion particles are identified with some squarks $\widetilde{q}$ (eventually parametrizing a standard model $D$-flat direction), they carry a $U(1)$ baryonic global charge which is derived from the conserved current
\begin{equation}
J^\mu_B=\frac{i}{2}\sum_q  q_B^q\left(\widetilde{q}^\dagger\partial^\mu\widetilde{q}-\widetilde{q}\partial^\mu\widetilde{q}^\dagger\right),
\end{equation}
where $q_B^q$ is the baryonic charge associated to any field $\widetilde{q}$. Under the assumption of cylindrical symmetry, the baryonic charge per unit length $Q_B$ along the $z$-axis will be given by $Q_B=\int d\theta dr\:r j_B(\theta,r)$ where $j_B$ is the current per unit length along the same axis.

Are the superconducting strings stable?
If there is no quantum tunnel transition on the string such that $\widetilde{f}$ passes through zero, $N$ is conserved and the string is stable. However, if the current on the string gets larger and larger because of some unspecified mechanism, the probability of quantum jumps increases. This results in a maximum  value for the current,  $I_{{\rm max}}\sim e\sqrt{\xi}/g\sim 10^{10} $ A \cite{superc}.

{\bf 4.}~~ Here we speculate as to the possible implications of the superconducting strings. As the temperature of the Universe decreases and the latter goes through the phase transition at the critical  temperature $T_c\sim\sqrt{\xi}$, some of the string-like configurations that were undergoing thermal fluctuations will freeze out and thus not decay.
After the phase transition, topological strings can only occur as closed loops or infinite strings. 
 
Because strings are superconducting, during the transition random currents will be induced on the strings. The net current on the loop of a size $\ell$ is expected to be proportional to $T_c(\ell/\xi_I)^{1/2}$ where $\xi_I\sim T_c^{-1}$ is the correlation length of the random currents. 
The dynamics of the loops is governed by the tension of the string, frictional forces and Hubble expansion. Initially the frictional forces are very large and so the string motion is highly damped. The friction dominated period lasts until the temperature of the Universe has dropped down to the value $T_*\simeq 
(\xi/\Mpl)$. Since we are interested in low values of supersymmetry breaking scale, only the friction dominated period of string evolution will be relevant. 
During this era loops would tend to collapse under their own tension. However, the effective tension of the string is the sum of the bare tension and the square of the current on the string \cite{n1,n2}. 
As the loops collapses, the current builds up and the effective tension becomes smaller. In such a situation two possibilities may occur: either the  current may become so   large that the charge carriers can leave the string, that is, the current can saturate, or the effective action goes to zero and the loop does not collapse any further, forming a vorton. If the first possibility is realized, the loops  continue to collapse and eventually disappear.  If the second possibility is realized, the loops form static ring configuration that can survive until some quantum tunneling causes the charge to leak. Sfermion excitations  may leave the world-sheet if their mass in the core of the string is smaller compared to the energy scale of the underlying vortex, though some  angular momentum when both $N$ and the non-topological charges are present can help stabilizing the string. The decay rate is estimated to be of the order of ${\rm exp}\left[-N\left(g^2/4\pi\right)^{3/2}\right]$ \cite{n2}, so that the bosonic vo
rtices are expected to be long-lived for total winding numbers $N\gg  \left(g^2/4\pi\right)^{-3/2}$. 

In both cases, depending on the lifetime of the loops and their decay products, their cosmology may be of great interest, especially if  a fixed amount of baryonic charge  is stored in the string. The generation of the baryon number might occur in some novel way when vortices finally collapse and disappear.  Another conceivable scenario is when  two straight superconducting strings come into contact. They will start to align because of the strong magnetic interaction of their currents. If one of the string forms a loop, the other would wind up to this loop and the electric field during this gluing up   might be large enough to copiously produce pairs of heavy particles and produce the  baryon asymmetry \cite{rep}.  

With regard to the issue of baryon asymmetry generation, it is important to recall that the presence of the superconducting string is expected to result in the electroweak symmetry being "restored" in the region surrounding the string\footnote{What one really means by symmetry restoration is that 
the standard model Higgs field configurations may be driven to zero   in the core of the strings. The standard model gauge group is usually already broken by the sfermion condensate.}
 out to a typical distance of the order of $r\sim 10^{-5} I\:{\rm GeV}^{-2}$, where $I$ is the current along the string \cite{davis}. This happens because the sfermion excited states on the string  carry hypercharge  and possibly  $T_3$ charge and the physical origin of the symmetry restoration can be directly traced to the coupling of the $Z$-field and the Higgs bosons to the strings. 

It is interesting to notice that these low energy SUSY breaking  local strings will carry $Z$-flux inside their core.   Moreover, $W$-condensation may take place in some region close to the string \cite{davis}. If the current inside the string is large enough, the typical length scale of the region where the Higgs field configurations vanish may be larger than $\sim M_W^{-1}$ and sphaleron type effects might  be unsuppressed. In such a case, the baryon and the lepton numbers are violated by quantum effects inside the string, but still the linear combination $B-L$ is conserved. This means that if a superconducting string with a certain amount of lepton number is generated and at some point of its evolution the current $I$ is large enough, it   
might  be accompanied by the generation of some baryonic charge. Sphalerons are not always active in the string though. Indeed, {\it any} condensate $\phi_i$ carrying
$SU(2)_L$ quantum numbers contribute to the sphaleron energy, $E_{{\rm sph}}\propto\sqrt{\sum_i|\phi_i|^2}$. Therefore, if the sfermion condensate in the core of the string is charged under $SU(2)_L$, sphaleron transitions are expected to be suppressed even though the Higgs field configurations vanish in the vortex.

Superconducting strings  may also be relevant for the generation of a primordial  magnetic field \cite{s1,s2}. Small seed fields may be produced by the network of charged-current carrying strings and subsequently amplified by the dynamo mechanism to generate the  magnetic field strength observed in our galaxy and many other spiral galaxies \cite{k}. This phenomenon depends upon the importance of equilibration processes during the evolution of the currents  \cite{s1}. Superconducting strings formed at the electroweak scale seem to give rise to magnetic fields smaller than required. However, the typical energy scale of the superconducting strings predicted in the framework of low energy SUSY breaking models is appreciably larger than the weak scale and a viable seed magnetic field for the galactic dynamo could be generated if the equilibration is weak. This and other issues , such as the generation of the baryon asymmetry, are presently under investigation.

\vskip 1cm
\centerline{\bf Acknowledgements}

The author would like to thank G. Dvali for many useful comments and suggestions  and for a careful reading of the manuscript. C. Martins,  S. Trivedi and X. Zhang are also acknowledged for discussions. 
A.R. supported by the DOE and NASA under Grant NAG5--2788.

\def\NPB#1#2#3{Nucl. Phys. {\bf B#1}, #3 (19#2)}
\def\PLB#1#2#3{Phys. Lett. {\bf B#1}, #3 (19#2) }
\def\PLBold#1#2#3{Phys. Lett. {\bf#1B} (19#2) #3}
\def\PRD#1#2#3{Phys. Rev. {\bf D#1}, #3 (19#2) }
\def\PRL#1#2#3{Phys. Rev. Lett. {\bf#1} (19#2) #3}
\def\PRT#1#2#3{Phys. Rep. {\bf#1} (19#2) #3}
\def\ARAA#1#2#3{Ann. Rev. Astron. Astrophys. {\bf#1} (19#2) #3}
\def\ARNP#1#2#3{Ann. Rev. Nucl. Part. Sci. {\bf#1} (19#2) #3}
\def\MPL#1#2#3{Mod. Phys. Lett. {\bf #1} (19#2) #3}
\def\ZPC#1#2#3{Zeit. f\"ur Physik {\bf C#1} (19#2) #3}
\def\APJ#1#2#3{Ap. J. {\bf #1} (19#2) #3}
\def\AP#1#2#3{{Ann. Phys. } {\bf #1} (19#2) #3}
\def\RMP#1#2#3{{Rev. Mod. Phys. } {\bf #1} (19#2) #3}
\def\CMP#1#2#3{{Comm. Math. Phys. } {\bf #1} (19#2) #3}

\end{document}